\documentclass[
  aps,prb,twocolumn,showpacs,amsmath,amssymb,superscriptaddress
]{revtex4-1}

\makeatletter
\newcommand{\whencolumns}[2]{\preprintsty@sw{#1}{#2}}
\makeatother

\usepackage[english]{babel}
\usepackage{amsmath,amssymb,amsfonts} 	% Typical maths resource package
\usepackage{graphicx}
\usepackage{array,multirow}
\usepackage[utf8]{inputenc}
\usepackage{bm}
\usepackage{xcolor}
\usepackage[normalem]{ulem}
\usepackage{siunitx}
\usepackage{hyperref}
\hypersetup{colorlinks=true,linkcolor=blue,citecolor=blue,urlcolor=blue}
\usepackage{url}
\usepackage{textcomp}
\usepackage{lineno,hyperref}
\usepackage{float}
\usepackage{textcomp} %for n^o
\usepackage{comment}
\usepackage{rotating}
\usepackage[figurename=FIG.]{caption}
\usepackage{subcaption}
\usepackage{color} % only to put comments in the draft
\definecolor{red}{rgb}{1.,0.,0.}
\usepackage{soul}
\usepackage{amsmath}
\usepackage{enumitem}

 \usepackage[
 justification=raggedright]
 {caption}
\usepackage{subcaption}
\usepackage{qcircuit}
\usepackage{braket}

\makeatletter
\newcommand*{\rom}[1]{\expandafter\@slowromancap\romannumeral #1@}
\makeatother

\usepackage{siunitx}
\usepackage{chemformula}
\usepackage{xspace}
 
\newcommand\Harvard{ John A. Paulson School of Engineering and Applied Sciences, Harvard University, Cambridge, MA 02138, USA}
\newcommand\Sapienza{Department of Physics, Sapienza University of Rome, Piazzale Aldo Moro 5, 00185 Rome, Italy }
\newcommand\Bosch{Robert Bosch LLC Research and Technology Center, Watertown, MA 02472, USA}

\DeclareMathAlphabet\mathbfcal{OMS}{cmsy}{b}{n}

\newcommand{\mR}{\mathbfcal{R}}
\newcommand{\mP}{\mathbfcal{P}}
\newcommand{\bA}{\mathbf{A}}
\newcommand{\bB}{\mathbf{B}}
\newcommand{\bG}{\bm{\Gamma}}
\newcommand{\bR}{\mathbf{R}}
\newcommand{\bP}{\mathbf{P}}
\newcommand{\af}{\langle\mathbf{f}\rangle}
\newcommand{\dV}{\langle\bm{\partial}^2V\rangle}

\newcommand{\by}{\mathbf{y}}
\newcommand{\bby}{\mathbf{\bar{y}}}

\newcommand{\secname}{Sec.\xspace}
\newcommand{\eqname}{Eq.\xspace}
\begin{document}

\title{Atomistic simulations of out-of-equilibrium quantum nuclear dynamics}

\author{Francesco Libbi}
%\email[Corresponding author. ]{libbi@g.harvard.edu}
\thanks{Corresponding authors: \textcolor{blue}{libbi@g.harvard.edu}, \textcolor{blue}{lorenzo.monacelli@uniroma1.it}, \textcolor{blue}{bkoz@g.harvard.edu}}
\affiliation{\Harvard}
\author{Anders Johansson}
\affiliation{\Harvard}
\author{Lorenzo Monacelli}
%\email[Corresponding author. ]{lorenzo.monacelli@uniroma1.it}
\thanks{Corresponding authors: \textcolor{blue}{libbi@g.harvard.edu}, \textcolor{blue}{lorenzo.monacelli@uniroma1.it}, \textcolor{blue}{bkoz@g.harvard.edu}}
\affiliation{\Sapienza}
\author{Boris Kozinsky}
%\email[Corresponding author. ]{bkoz@g.harvard.edu}
\thanks{Corresponding authors: \textcolor{blue}{libbi@g.harvard.edu}, \textcolor{blue}{lorenzo.monacelli@uniroma1.it}, \textcolor{blue}{bkoz@g.harvard.edu}}
\affiliation{\Harvard}
\affiliation{\Bosch}

\begin{abstract}
The rapid advancements in ultrafast laser technology have paved the way for pumping and probing the out-of-equilibrium dynamics of nuclei in crystals. However, interpreting these experiments is extremely challenging due to the complex nonlinear responses in systems where lattice excitations interact, particularly in crystals composed of light atoms or at low temperatures where the quantum nature of ions becomes significant.
In this work, we address the nonequilibrium quantum ionic dynamics from first principles. Our approach is general and can be applied to simulate any crystal, in combination with a first-principles treatment of electrons or external machine-learning potentials. It is implemented by leveraging the nonequilibrium time-dependent self-consistent harmonic approximation (TD-SCHA), with a stable, energy-conserving, correlated stochastic integration scheme that achieves an accuracy of $\mathcal{O}(dt^3)$.
We benchmark the method with both a simple one-dimensional model to test its accuracy and a realistic 40-atom cell of \ch{SrTiO3} under THz laser pump, paving the way for simulations of ultrafast THz-Xray pump-probe spectroscopy like those performed in synchrotron facilities.

\end{abstract}

\maketitle

\section{Introduction}
Nuclear quantum effects often play a crucial role in determining properties of materials\cite{Markland2018}, affecting their thermodynamic stability \cite{Errea2016, Errea2020, PhysRevLett.96.227602}, electronic structure \cite{PhysRevLett.115.177401, PhysRevLett.112.215501}, and transport phenomena \cite{PhysRevLett.122.075901}. Understanding and accurately accounting for quantum contributions in tunneling and vibrational statistics requires simulations that go beyond classical nuclei approximation. Path-integral molecular dynamics (PIMD) is the most common approach to simulate nuclear quantum effects in complex anharmonic crystals \cite{10.1063/1.471221, RevModPhys.67.279, 10.1063/1.446740, 10.1063/1.441588}, especially with the ring-polymer formulation \cite{10.1063/1.1777575, 10.1063/1.2074967}, which is an exact framework for sampling the equilibrium nuclear density matrix. 
However, PIMD is rigorously formulated with an assumption of thermodynamic equilibrium as it evolves the trajectories in imaginary time. Simulating out-of-equilibrium quantum nuclear dynamics requires a theory for the real-time evolution, such as the path-integral quantum Monte Carlo (PIMC),  \cite{10.1063/1.472798,10.1063/1.473231, 10.1063/1.480028,10.1063/1.1703704}, which is extremely challenging due to the appearance of the so-called sign problem, in which different trajectories contribute with different signs, introducing noise in determining dynamical observables \cite{Alexandru_2022}. Therefore, most applications of real-time PIMC remain limited to systems composed of only a few degrees of freedom \cite{PhysRevLett.100.176403}.

Several techniques have been proposed to overcome the limits of real-time PIMC to study nuclear quantum effects in systems of realistic interest.
Among these, the time-dependent self-consistent harmonic approximation (TD-SCHA\cite{PhysRevB.103.104305,PhysRevB.107.174307}) holds significant promise.
The theory extends the stochastic self-consistent harmonic approximation (SSCHA)\cite{Monacelli_2021, PhysRevB.96.014111, PhysRevB.98.024106},  a well-established technique to simulate equilibrium thermodynamics of solids accounting for quantum nuclear fluctuations. The success of SSCHA lies in the adoption of approximations that are particularly effective for crystals, achieving a computational cost that is orders of magnitude lower than PIMD, while still producing predictions in very good agreement with experiments \cite{Errea2016, Errea2020, PhysRevLett.122.075901, PhysRevLett.125.106101, Aseginolaza2024}. 

TD-SCHA has already been employed in the linear response regime, where it enabled the prediction of Raman and IR spectra of metallic hydrogen with unprecedented accuracy \cite{Monacelli_Nature_2021}. However, up to now, no application of TD-SCHA beyond equilibrium has been attempted. The difficulty in obtaining accurate and stable dynamic solutions of TD-SCHA equations is challenging due to the need for the evaluation of ensemble averages of the nuclear potential energy landscape. 

In this work we address this challenge by developing an algorithm to solve the dynamical TD-SCHA equations and simulate the out-of-equilibrium dynamics of nuclei in complex realistic systems. In \secname~\ref{theory}, we revise the TD-SCHA equations of motion for the nuclear density matrix. \secname~\ref{FD} introduces three different numerical algorithms to integrate the TD-SCHA equations, and their stability is discussed in \secname~\ref{stability}. 
\secname~\ref{stochastic} presents the correlated sampling technique to perform the evaluation of the stochastic quantum averages. This formulation ensures both the efficient evaluation of ensemble averages and the numerical stability of the equations. Crucially, we show that such a correlated approach conserves energy in one dimensional problems independently of the number of stochastic configurations adopted.  We benchmark the method in \secname~\ref{tests}, where we compare the different numerical schemes on a one dimensional model system. Finally, in Sec. \ref{STO}, we provide an example of the application of TD-SCHA to realistic systems by studying the quantum dynamics in $\mathrm{SrTiO_3}$ (STO) when driven out of equilibrium by a strong laser pulse of THz frequency.

\section{Time-dependent self-consistent harmonic approximation}\label{theory}

The TD-SCHA formulation leverages the Wigner formalism\cite{PhysRev.40.749,10.1063/1.1705323}.
The Wigner transform of the nuclear quantum density matrix $\hat{\rho}(t)$ describing the quantum state is defined as 
\begin{equation}
    \rho(\bR, \bP, t) = \int \frac{ e^{-\frac{i}{\hbar}\mathbf{P}\cdot\bR'}}{(2\pi\hbar)^{3N}} \Bigl\langle \bR + \frac{\bR'}{2}\Bigl|\hat{\rho}(t) \Bigr| \bR - \frac{\bR'}{2}
    \Bigr\rangle  d\bR'\ ,
\end{equation}
and it maps the quantum operator $\hat{\rho}(t)$ into a function of positions and momenta $\rho(\bR, \bP, t)$ that is analogous to the classical nuclear density. All atomic quantities are rescaled by mass to simplify the notation: $R_{i} = \tilde R_{i}\sqrt{m_i}$ and $P_{i} = \tilde P_{i}/\sqrt{m_i}$, where the index $i$ goes from 1 to $3N$, containing both the Cartesian and atomic index, and the tilde indicates the standard (not mass-rescaled) quantities. Dynamical averages of any quantum observable are obtained by tracing $\bR$ and $\bP$ on the density:
\begin{equation}\label{avgO}
    \left<O\right>(t) = \int d\bR d\bP \rho(\bR, \bP, t) O(\bR, \bP).
\end{equation}
The TD-SCHA method is based on expressing the Wigner density matrix as a general Gaussian form in terms of ionic positions $\bR$ and momenta $\bP$:
\begin{equation}\label{density}
     \rho(\bR, \bP, t) = \frac{1}{\mathcal{N}}e^{-\frac{1}{2}( {\delta\bR}\cdot\bm{\alpha}\cdot {\delta\bR} +  {\delta\mathbf{P}}\cdot\bm{\beta}\cdot {\delta\mathbf{P}} +  {\delta\bR}\cdot\bm{\gamma}\cdot {\delta\mathbf{P}} )}\ .
\end{equation}
 Here $ {\delta\bR}(t)=  \bR-\mR(t)$ and $ {\delta\mathbf{P}}(t)= \bP-\mP(t)$ and $\mathcal{N}$ is the normalization factor, where $\mR(t) = \braket{\bR}(t)$ represent average positions and $\mP(t) = \braket{\bP}(t)$ the average momenta.
 The $\bm{\alpha}(t)$, $\bm{\beta}(t)$ and $\bm{\gamma}(t)$ matrices are related to, respectively, position-position, momentum-momentum, and position-momentum covariances by the following relations:
\begin{equation}
    \bA^{-1} = \braket{\delta R_i \delta R_j}^{-1}= \bm{\alpha}-\bm{\gamma}\cdot\bm{\beta}^{-1}\cdot\bm{\gamma}^T\ ,
\end{equation}
\begin{equation}
    \bB^{-1} = \braket{\delta P_i \delta P_j}^{-1} = -\bm{\gamma}^T +\bm{\beta}\cdot\bm{\gamma}^{-1}\cdot\bm{\alpha}\ ,
\end{equation}
\begin{equation}
    \bG^{-1} = \braket{\delta R_i \delta P_j}^{-1} = \bm{\beta}-\bm{\gamma}^T\cdot\bm{\alpha}^{-1}\cdot\bm{\gamma}^T\ .
\end{equation}
The evolution of the density Wigner-space density matrix $\rho(\bR, \bP, t)$ is determined by the propagation in time of the parameters $\mR(t)$, $\mP(t)$, $\bA(t)$, $\bB(t)$ and $\bG(t)$. 
Analogously to the time-dependent Hartree-Fock or time-dependent density functional theory for electrons, the time evolution is obtained by imposing the least action principle \cite{PhysRevB.107.174307}, leading to the time-dependent self-consistent Liouville-von Neumann equation for the density matrix:
\begin{equation}\label{liouville}
    i\hbar\frac{\partial\hat\rho}{\partial t} = \left[\mathcal H[\hat\rho], \hat\rho\right],
\end{equation}
where $\mathcal H[\hat \rho]$ is a self-consistent harmonic Hamiltonian whose parameters depend on the anharmonic potential and the density matrix $\hat\rho$ and the square brackets indicate the quantum commutator (more details in Ref.\cite{PhysRevB.103.104305}). Notably, in TD-SCHA, $\mathcal H[\hat \rho]$ is local in time, so the time evolution depends only on the current quantum state.
Expressing Eq. \ref{liouville} in the Wigner formalism and substituting the Gaussian form for the density matrix leads to the set of differential equations
\begin{equation}\label{tdscha}
\begin{cases}
    \dot{\mR} = \mP \\
    \dot{\mP} = \braket{\mathbf{f}} \\
    \dot{\bA} = \bG + \bG^{\dag} \\
    \dot{\bB} =  -\dV\bG - \bG^{\dag}\dV  \\
    \dot{\bG} =   \bB-\bA\dV 
\end{cases} \ ,
\end{equation}
where the dot over a tensor $\dot{\circ}$ indicates the time-derivative, the product between tensors is the standard rows-by-columns contraction among all indices, and the dagger symbol indicates the matrix transposition operation $O^\dag_{ij} = O_{ji}$.  
Here, the atomic potential energy landscape (PES), $V(\bR, t)$, enters the quantum averages of forces $\af$ and the average curvature tensor $\dV$, defined as
\begin{equation}
    \langle f_{a}\rangle(t) = -\int d\bR d\bP \frac{\partial V}{\partial R_{a}}(t) \rho(\bR, \bP, t),
    \label{eq:force}
\end{equation}
\begin{equation}
    \langle\partial_{ab}^2V\rangle(t) = \int d\bR d\bP \frac{\partial^2 V}{\partial R_{a} \partial R_{b}}(t)\rho(\bR, \bP, t).
    \label{eq:d2v}
\end{equation}
The solution of \eqname~\eqref{tdscha} provides the quantum state $\rho(\bR, \bP, t)$, enabling the direct computation of the time envelope of any quantum observable.\\
The stationary solution of \eqname~\eqref{tdscha} coincides with the equilibrium fixed-volume state that minimizes the Helmholtz free energy \cite{} and can be obtained with the standard SSCHA algorithm\cite{PhysRevB.107.174307,Monacelli_2021}. \\
When simulating a pump-probe experiment, the system is prepared at equilibrium and perturbed with a radiation pulse modeled as a time-dependent external potential $V_\text{ext}(\bR, t)$. The overall potential that enters in \eqname~\eqref{eq:force} and \eqref{eq:d2v} is 
\begin{equation}
    V(\bR, t) = V_\text{BO}(\bR) + V_\text{ext}(\bR, t),
\end{equation}
where $V_\text{BO}(\bR)$ is the instantaneous interaction potential of nuclei within the Born-Oppenheimer approximation that depends only on the nuclear positions.

\section{Numerical integration of TD-SCHA equations}\label{FD}

Numerical solutions of the TD-SCHA equations of motion \eqref{tdscha} have so far been limited to simple one-dimensional models \cite{PhysRevB.107.174307} and linear-response calculations, enabled by an efficient Lanczos algorithm\cite{PhysRevB.103.104305}. The major challenge to applying TD-SCHA in the out-of-equilibrium regime is associated with the cost of calculating the averages $\af$ and particularly $\dV$,  \eqname~\eqref{eq:d2v}, as sampling the second derivatives of the potential is computationally expensive. 
This section introduces a finite-difference scheme to integrate the TD-SCHA equations with an error scaling as $\mathcal{O}(dt^3)$ that requires the computation of \eqname~\eqref{eq:d2v} only once per time step. 
Expanding the time evolution of TD-SCHA quantities in the Taylor series to  second order in the time step increment $dt$, we derive the following expressions:

\begin{widetext}
\begin{equation}\label{tdscha_taylor}
\begin{cases}
    \mR_{t+dt} = \mR_t + \mP_t dt + \frac{1}{2}\braket{\mathbf{f}_t}dt^2 + \mathcal{O}(dt^3) \\
    \mP_{t+dt} = \mP_t + \braket{\mathbf{f}_t}dt + \frac{1}{2}\mP''_tdt^2 + \mathcal{O}(dt^3) \\
    \bA_{t+dt} = \bA_t + \Bigl(\bG + \bG^{\dag}\Bigr)_tdt + 
    \frac{1}{2}\Bigl(\bB-\bA\dV\Bigr)_t dt^2 + 
    \frac{1}{2}\Bigl(\bB-\dV \bA\Bigr)_t dt^2 + \mathcal{O}(dt^3)\\
    \bB_{t+dt} = \bB_t - \Bigl(\dV\bG + \bG^{\dag}\dV\Bigr)_tdt + \frac{1}{2}\bB''_tdt^2  + \mathcal{O}(dt^3)\\
    \bG_{t+dt} = \bG_t + \Bigl(\bB-\bA\dV \Bigr)_t dt + \frac{1}{2}\bG''_t dt^2 + \mathcal{O}(dt^3)
\end{cases}
\end{equation}
\end{widetext}

\noindent

Notably, while the equations for the evolution of $\mR$ and $\bA$ are explicit up to $\mathcal{O}(dt^3)$, we need the values of the second derivatives of $\bB_t$, $\bG_t$ and $\mP_t$.

To preserve the $\mathcal{O}(dt^3)$ error of the time propagation we use the central difference formula to approximate those second derivatives (see Appendix \ref{integr}):
\begin{equation}\label{mixed}
F_{t+dt} = F_t + \frac{1}{2}(F'_t+F'_{t+dt})dt + \mathcal{O}(dt^3)\ ,
\end{equation}
where $F$ represents a generic variable. This expression only requires the knowledge of its first derivatives at times $t$ and $t+dt$, but not of the second derivative.
We can rely on two observations:
    (i) The calculation of $\af$ and $\dV$ depends only on  $\mR$ and $\bA$ as the potential is a function of only the positions (see Eqs.~\ref{eq:force}-\ref{eq:d2v} and Appendix \ref{wigner});
    (ii) both $\mR$ and $\bA$ can be integrated explicitly with accuracy $\mathcal{O}(dt^3)$.
The GV algorithm we devise comprises the following steps:
(I) calculate $\mR_{t+dt}$ and $\bA_{t+dt}$ with accuracy $\mathcal{O}(dt^3)$ using the first and the third of Eqs. \ref{tdscha_taylor} respectively;
(II) use $\mR_{t+dt}$ and $\bA_{t+dt}$ to determine $\braket{\mathbf{f}}_{t+dt}$ and ${{\dV}_{t+dt}}$;
(III) determine $\mP$, $\bB$ and $\bG$ using  Eq. \ref{mixed}
    \begin{equation}\label{last}
    \begin{cases}
        \mP_{t+dt} = \mP_t + 
        \af_{\bar t}\,dt 
        + \mathcal{O}(t^3)   \\
        \bB_{t+dt} = \bB_t - \Bigl(\dV\bG + \bG^{\dag}\dV\Bigr)_{\bar{t}}dt+  \mathcal{O}(dt^3)\\
        \bG_{t+dt} = \bG_t + \Bigl(\bB-\bA\dV\Bigr)_{\bar{t}}dt +  \mathcal{O}(dt^3)\ .
    \end{cases}
\end{equation}
Here we use the shorthand $F_{\bar{t}}$ to indicate
$\frac{1}{2}(F_t+F_{t+dt})$.
Even though \eqname~\ref{last} is implicit in the variables $\bB$ and $\bG$, it requires the calculation of $\af$ and $\dV$ just once per time step (due to observation (i)), and it evolves the parameters with accuracy $\mathcal{O}(dt^3)$. Furthermore, the equations for integrating $\mR$ and $\mP$ coincide with the familiar \textit{velocity Verlet} scheme \cite{PhysRev.159.98}.

We now compare the above GV integration scheme with alternatives, namely explicit Euler (EE) and semi-implicit Euler (SIE) algorithms, showing that the GV is more accurate than both EE and SIE and stable for larger values of the time-step $dt$.
Alternative integration strategies for Eqs.~\eqref{tdscha_taylor} are founded on the semi-implicit Euler method. Since $\bG$ is related to the derivatives of $\bA$ and $\bB$ (see Eqs.~\ref{tdscha}), the semi-implicit Euler (SIE) scheme consists in updating $\bG$ first, then computing $\bA_{t+dt}$ and $\bB_{t+dt}$ by using $\bG_{t+dt}$ instead of $\bG_t$:
\begin{equation}
\begin{cases}
    \bA_{t+dt} = \bA_t + \Bigl(\bG + \bG^{\dag}\Bigr)_{t+dt}dt + \mathcal{O}(dt^2)  \\
    \bB_{t+dt} = \bB_t - \Bigl(\dV_t\bG_{t+dt} + \bG^{\dag}_{t+dt}\dV_t\Bigr)dt + \mathcal{O}(dt^2)  \\
    \bG_{t+dt} = \bG_t + \Bigl(\bB-\bA\dV \Bigr)_t dt  + \mathcal{O}(dt^2) \\
\end{cases}\ .
\end{equation}
Instead, $\mR$ and $\mP$ are evolved according to Verlet. 
These integration schemes can be compared to the simplest approach, the explicit Euler (EE) scheme, where all the parameters are evolved simultaneously:
\begin{equation}
\begin{cases}
    \bA_{t+dt} = \bA_t + \Bigl(\bG + \bG^{\dag}\Bigr)_{t}dt + \mathcal{O}(dt^2)  \\
    \bB_{t+dt} = \bB_t - \Bigl(\dV_t\bG_{t} + \bG^{\dag}_{t}\dV_t\Bigr)dt + \mathcal{O}(dt^2) \\
    \bG_{t+dt} = \bG_t + \Bigl(\bB-\bA\dV \Bigr)_t dt  + \mathcal{O}(dt^2)  \\
\end{cases}\ .
\end{equation}

\section{Stability of the integration schemes}\label{stability}
Here, we investigate the stability of the integration schemes introduced in the previous section. Particularly, we are interested in the dynamics of the variables $\bA$, $\bB$, and $\bG$, which are not present in classical Newtonian nuclear dynamics equations. 

Let us consider a 1D wave packet evolving in a Harmonic potential. 
Thanks to the constant curvature $\bm{\kappa}$ of the PES, $\bA$, $\bB$ and $\bG$ do not depend on the average coordinates (centroids) $\mR$ and $\mP$ since
\begin{equation}
    \dV = \bm{\kappa}.
\end{equation}
The TD-SCHA equations for these variables reduce to
\begin{equation}
    \begin{pmatrix}
        \dot{A}\\\dot{B}\\\dot{\Gamma}
    \end{pmatrix}
    = 
    \begin{pmatrix}
        0 & 0 & 2\\
        0 & 0 & -2\kappa\\
        -\kappa & 1 &0 
    \end{pmatrix}
    \begin{pmatrix}
        A \\ B \\ \Gamma
    \end{pmatrix}\ . 
\end{equation}
By rescaling the parameters as
\begin{equation}\label{transf}
\begin{cases}
     A' = \sqrt{\frac{\kappa}{2}} A \\
     B' = \frac{1}{\sqrt{2\kappa}} B \\
     t' = \sqrt{2\kappa} t \\ 
\end{cases}
\end{equation}
we get a generalized equation that does not depend on the PES curvature $\kappa$ 
\begin{equation}
    \begin{pmatrix}
        \dot{A'}\\\dot{B'}\\\dot{\Gamma'}
    \end{pmatrix}
    = 
    \begin{pmatrix}
        0 & 0 & 1\\
        0 & 0 & -1\\
        -1 & 1 &0 
    \end{pmatrix}
    \begin{pmatrix}
        A' \\ B' \\ \Gamma'
    \end{pmatrix}\ .
\end{equation}
The above equation can be written in a compact notation as
\begin{equation}
    \dot{\mathbf{x}} = \mathbf{M} \cdot \mathbf{x}
\end{equation}
where the propagation matrix $M$ is skew-symmetric. This symmetry imposes that the norm of $\mathbf x$ is conserved
\begin{equation}
    A'^2 + B'^2 + \Gamma'^2 = \mathrm{const}\ . 
\end{equation}
In the following, we omit the prime symbol to maintain a cleaner notation and introduce the integer step $n$ as $n=t/dt$. The stability of the methods is investigated by calculating the step transformation matrix $\mathbf{S}(dt)$, which connects the degrees of freedom at time step $n+1$ with those at step $n$:
\begin{equation}\label{sn}
    \mathbf{x}_{n+1} = \mathbf{S}(dt) \mathbf{x}_n\ .
\end{equation}
Iterating \eqname~\eqref{sn}, we obtain
\begin{equation}
    \mathbf{x}_{n+1} = \mathbf{S}(dt)^n \mathbf{x}_0\ .
\end{equation}
The stability condition is achieved if the propagator $\mathbf{S}(dt)^n$ remains finite for arbitrary large powers $n$. This is equivalent to requiring that all its eigenvalues $\lambda$ are such that $|\lambda|\leq1$. 
We calculate the step transformation matrices for EE, SIE, and GV schemes. The details of the derivation are reported in Appendix \ref{integr}. 
For the EE method we find that 
\begin{equation}
    \lambda_{\mathrm{max}} = 1+2dt^2 > 1 \qquad  \forall dt\ ,   
\end{equation}
meaning that the EE method is \textit{unconditionally unstable}.
The stability condition obtained for the SIE method is instead
\begin{equation}
    dt_{SIE} \leq \frac{1}{\omega}\ ,
\end{equation}
where $\omega = \sqrt{\kappa}$ is the frequency of the harmonic oscillator, while for the GV method we obtain
\begin{equation}
    dt_{GV} \leq \frac{\sqrt{2}}{\omega}\ .
\end{equation}
Thus, both the SIE and GV are stable for sufficiently small $dt$, with the GV method having a larger stability range. The integration step of the GV algorithm must thus be shorter than approximately 1/5th of the shortest period of vibrational motion of an atomic system.

\section{Stochastic formulation}\label{stochastic}
The ensemble averages of the potential energy and its derivatives are multidimensional integrals that are challenging to calculate. One strategy for addressing this issue involves expanding the potential energy in a Taylor series centered at a high-symmetry point of the structure , which allows analytically computing the thermodynamic averages. This approach relies on the analytic knowledge of Gaussian integrals, and it is at the basis of the self-consistent phonon (SCP) approach\cite{ALAMODE}.
The alternative consists in evaluating the integrals through a stochastic Monte Carlo algorithm, as exploited by the SSCHA approach \cite{Monacelli_2021}.
Here, we introduce a stochastic formulation for the TD-SCHA. The ensemble average of the potential energy on the nuclear density in \eqname~\eqref{density} can be calculated as:
\begin{equation}\label{vD}
    \braket{V}_{\mathcal{D}}=\frac{1}{N_c}\sum_{i=1}^{N_c} V(\mR(t) + \mathbf{J}(t)\cdot \by_i(t))\ ,
\end{equation}
where $\mathbf{y}^i(t)$ are i.i.d. normal random variables , $N_c$ is the number of stochastic configurations, and the subscript $\mathcal{D}$ stands for the discrete evaluation of the ensemble average (a more detailed introduction is in Appendix \ref{imp_samp}). $\mathbf{J}(t)$ is the principal square root (one of many possibilities) of the position-position covariance $\bA(t)$:
\begin{equation}\label{J}
    J_{ab} = \sum_{\mu} \sqrt{\lambda_{\mu}}e_{\mu a}e_{\mu b}\ ,
\end{equation}
where $\lambda_{\mu}$ and $e_{\mu}$ are respectively eigenvalues and eigenvectors of $\bA$.
The TD-SCHA equations only require the averages of the first and second derivatives of the potential. In our formulation, both averages require only the calculation of forces, which can be obtained either from first principles or from surrogate machine-learning force field models.
The average of the first derivative of the potential simplifies to 
\begin{equation}\label{fD}
    \Bigl\langle \frac{\partial V}{\partial R_a}\Bigr\rangle_{\mathcal{D}} = -\frac{1}{N_c}\sum_{i=1}^{N_c} f_a(\mR + \mathbf{J}\cdot\mathbf{y}_i) \ , 
\end{equation}
where the Cartesian force component is
\begin{equation}
    f_a = -\frac{\partial V}{\partial R_a}\ .
\end{equation}
(Here, we have omitted the time dependence for clarity in notation).
The calculation of the ensemble average of the second derivatives leverages integration by parts to solely utilize the forces \cite{PhysRevB.96.014111} 
\begin{equation}\label{kD}
    \Bigl\langle \frac{\partial^2 V}{\partial R_a \partial R_b}\Bigr\rangle_{\mathcal{D}} = -\sum_{cd} A^{-1}_{ac}\sum_{i=1}^{N_c} J_{cd} \ y_{d i} f_b(\mR + \mathbf{J}\cdot\mathbf{y}_i)\ . 
\end{equation}
Eq. \ref{kD} is symmetric in the Cartesian indexes $a$ and $b$ only in the limit $N_c\to\infty$. For a finite number of configurations, it must be symmetrized: 
\begin{equation}\label{kDs}
    \Bigl\langle \frac{\partial^2 V}{\partial R_a \partial R_b}\Bigr\rangle_{\mathcal{D}}^{sym} = 
    \frac{1}{2}\Bigl\langle \frac{\partial^2 V}{\partial R_a \partial R_b}\Bigr\rangle_{\mathcal{D}} + 
    \frac{1}{2}\Bigl\langle \frac{\partial^2 V}{\partial R_a \partial R_b}\Bigr\rangle_{\mathcal{D}}^T
\end{equation}
The stochastic evaluation of these integrals is characterized by Gaussian noise, which decreases as $1/N_c$. If the random displacements $\mathbf{y}_i$ are sampled in uncorrelated way at each time step, this implies the presence of stochastic noise as input in the TD-SCHA differential equations, which can significantly affect their accuracy and stability.
As demonstrated in Ref. \cite{PhysRevB.107.174307}, the TD-SCHA equations conserve the total energy in the absence of external potentials acting on the system:
\begin{equation}\label{dV}
\frac{d}{dt} \sum_{a}\Bigl[ \frac{B_{aa}+\mathcal{P}_a^2}{2} \Bigr] + \frac{d}{dt}\braket{V}=0
\end{equation}
However, the total energy is conserved only in the limit for $N_c\to\infty$. 
We can show that these issues can be fixed by using the same random configurations $\bby_i$ in different time step evaluations, which we refer to as correlated sampling. This choice introduces a systematic bias but eliminates the stochastic noise of the ensemble averages of forces and curvatures, making the time evolution smooth. Moreover, we can demonstrate that for one-dimensional problems, energy conservation holds true for any finite number of configurations when a constant $\bby_i$ is employed as $dt\to 0$ (see Appendix \ref{energy_conservation}). 
For higher-dimensional problems, energy conservation remains dependent on the number of configurations due to arbitrariness in the definition of $\mathbf{J}$. Nevertheless, the correlated sampling approach drastically improves the energy conservation for a given number of configurations $\mathrm{N}_c$, as shown numerically in the following sections. The possibility of choosing a gauge for $\mathbf{J}$ that allows for energy conservation independent of the number of configurations in higher-dimensional problems is discussed in the Appendix \ref{energy_conservation} and will be the subject of future research.

\begin{figure}[h]
    \centering
    \includegraphics[scale=0.48]{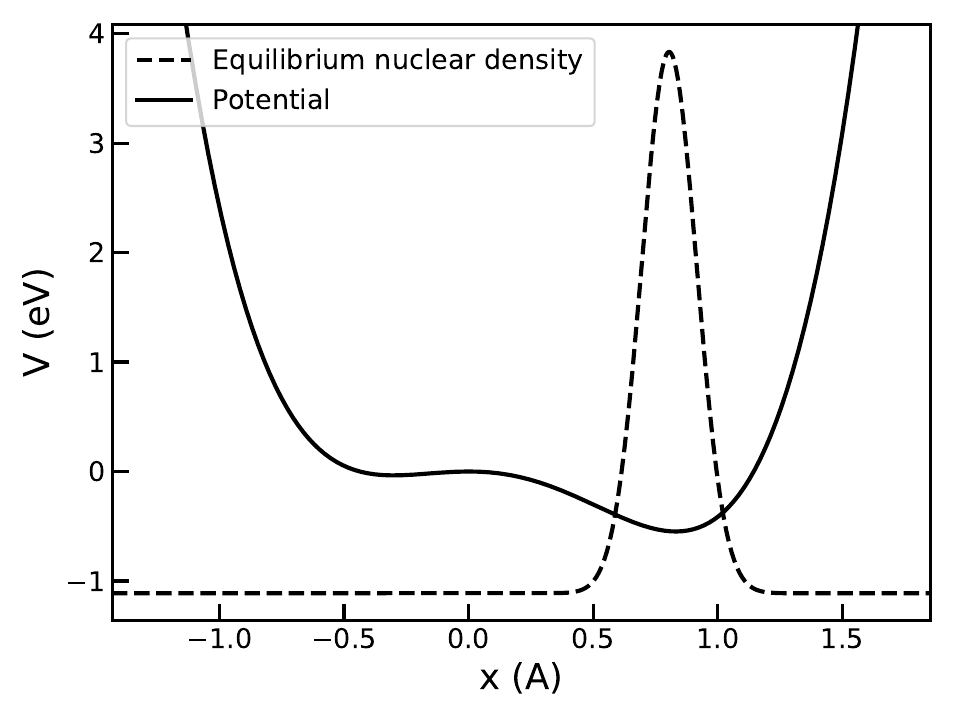}
    \caption{The solid curve represents the potential energy surface of the model, while the dashed Gaussian corresponds to the equilibrium nuclear distribution.}
    \label{pot}
\end{figure}

\begin{figure*}[t]
    \centering
        \includegraphics[scale=0.4]{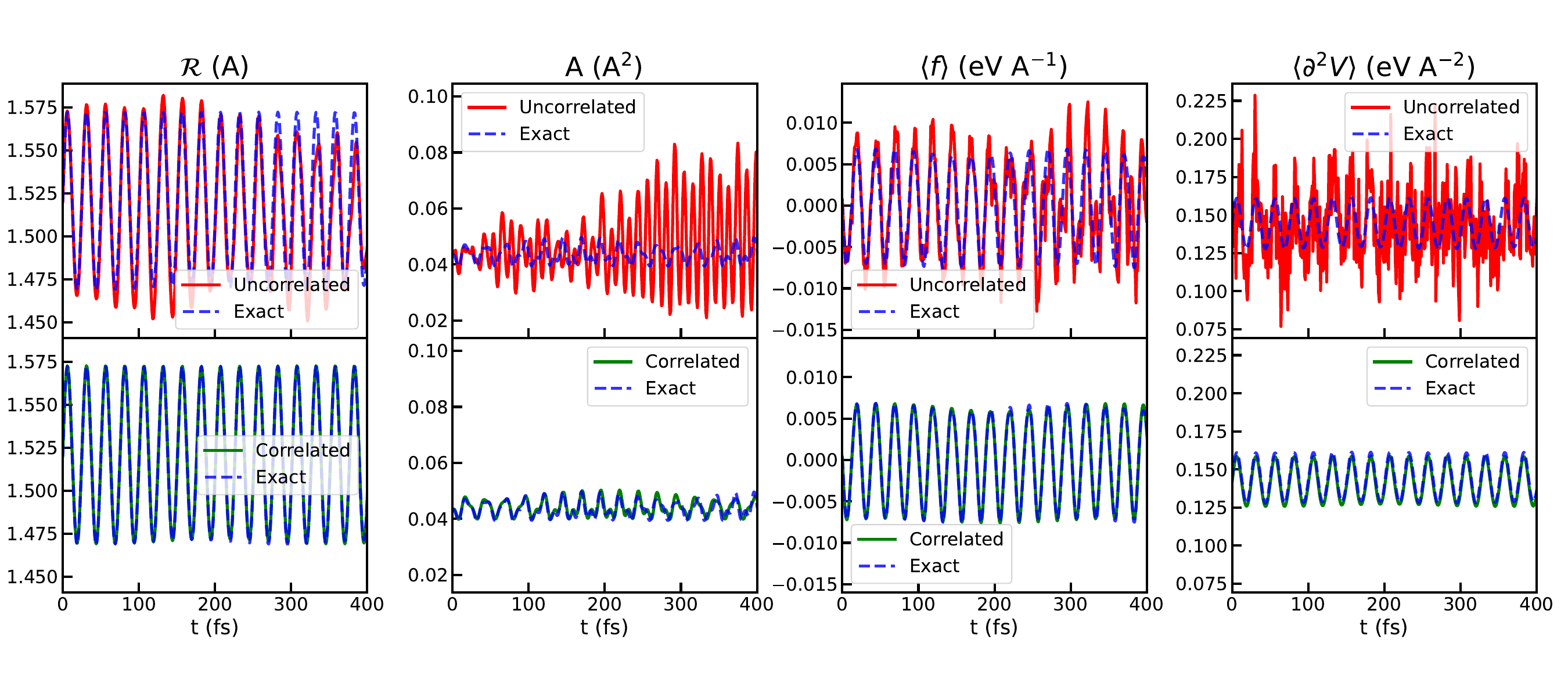}
    \caption{The panels in the upper row compare the $N_c\to\infty$ solution to the TD-SCHA equations (blue line) with the finite sampling solution using uncorrelated displacement (red line). The lower row of panels compares the exact solution to the TD-SCHA equation with the numerical solution using correlated sampling (green line). The quantities represented are (moving from left to right) the position $R$, position-position correlation $A$, average force $\braket{f}$, and average curvature $\braket{\partial^2V}$. For both the calculations with correlated and uncorrelated displacements, we employed $\mathrm{N_c}=100$ and a time-step of 1 fs.  }
    \label{total}
\end{figure*}

\section{Tests}\label{tests}
\begin{figure}
    \centering
    \includegraphics[scale=0.5]{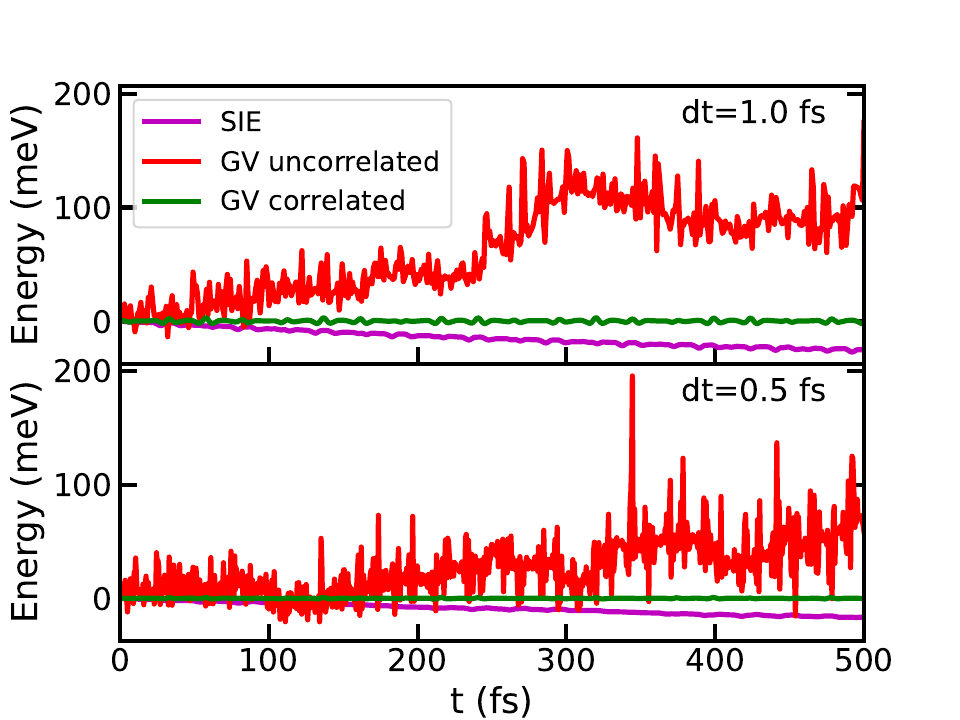}
    \caption{Energy conservation for different integration schemes and time-step. }
    \label{energy}
\end{figure}

We test the integration schemes on a one-dimensional model with potential energy 
\begin{equation}
    V(u) = \frac{1}{2}\Bigl(-au^2 - bu^3 + cu^4\Bigr) \ ,
\end{equation}
where $a=\SI{1.00}{\electronvolt\per\angstrom^2}$, $b=\SI{1.00}{\electronvolt\per\angstrom^3}$, and $c = \SI{1.00}{\electronvolt\per\angstrom^4}$.
The ionic mass is 1 amu. As illustrated in Fig. \ref{pot}, this potential exhibits two local minima separated by a barrier of approximately \SI{0.58}{\electronvolt}. The initial conditions for the parameters ${\mR, \bA, \bB, \bG}$ correspond to thermodynamic equilibrium, determined by solving the SSCHA equations \cite{Monacelli_2021} at \SI{100}{\kelvin}. In such equilibrium, the nuclear density is centered at the lowest minimum, with a spread of about 0.5\AA\ due to the ion's light mass. The initial momentum of the oscillator is set to $\mathrm{\frac{\mathcal{P}}{\sqrt{m}}=0.075\sqrt{eV}}$.
We solve the TD-SCHA equations using the stochastic formulation, comparing uncorrelated and correlated random displacements with $\mathrm{N_c}=100$ random configurations and a time-step of \SI{1}{\femto\second} for \SI{400}{\femto\second} with the GV scheme. The results, reported in \figurename~\ref{total}, are compared with the solution of the TD-SCHA equations in the $N_c\to\infty$ limit (numerically sampled via trapezoidal integration on a dense grid). The uncorrelated sampling algorithm quickly deviates from the $N_c\to\infty$ solution. The correlated sampling algorithm, instead, remains stable throughout the dynamics due to the suppression of the stochastic noise across different time steps.
\figurename~\ref{energy} report the energy conservation over time for the uncorrelated GV, correlated GV, and correlated SIE methods. The uncorrelated sampling approach fails to conserve energy, although reducing the simulation time step by half partially mitigates this issue. The SIE algorithm with correlated sampling suppresses the energy oscillations, but it suffers from a uniform energy drift that decreases with a reduction in the simulation time step. In contrast, the correlated GV method demonstrates flawless energy conservation for both time steps, exhibiting no energy fluctuations or drift. 
Indeed, the quality of energy conservation increases with the number of configurations for the uncorrelated sampling algorithm, while it is unaffected when employing the correlated sampling, as shown in \figurename~\ref{uncorr_ene} (see Appendix~\ref{energy_conservation} for the formal proof). %Note the  
\begin{figure}
    \centering
    \begin{subfigure}{0.48\textwidth}
        \includegraphics[scale=0.5]{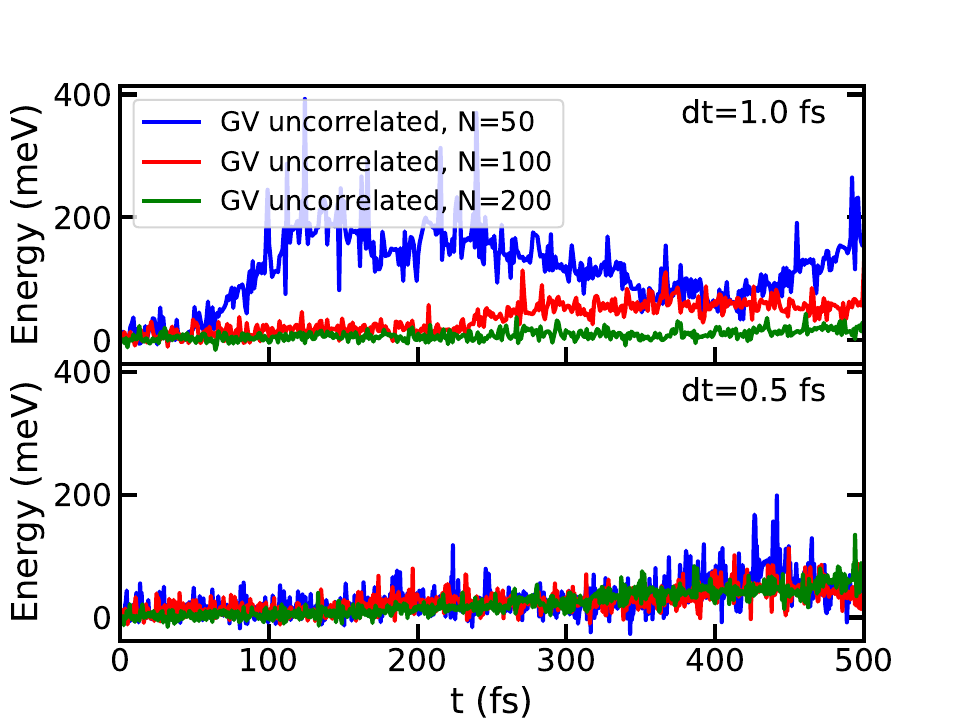}
        \caption{Uncorrelated}
        \label{uncorr_ene}
    \end{subfigure}
    \begin{subfigure}{0.48\textwidth}
        \includegraphics[scale=0.5]{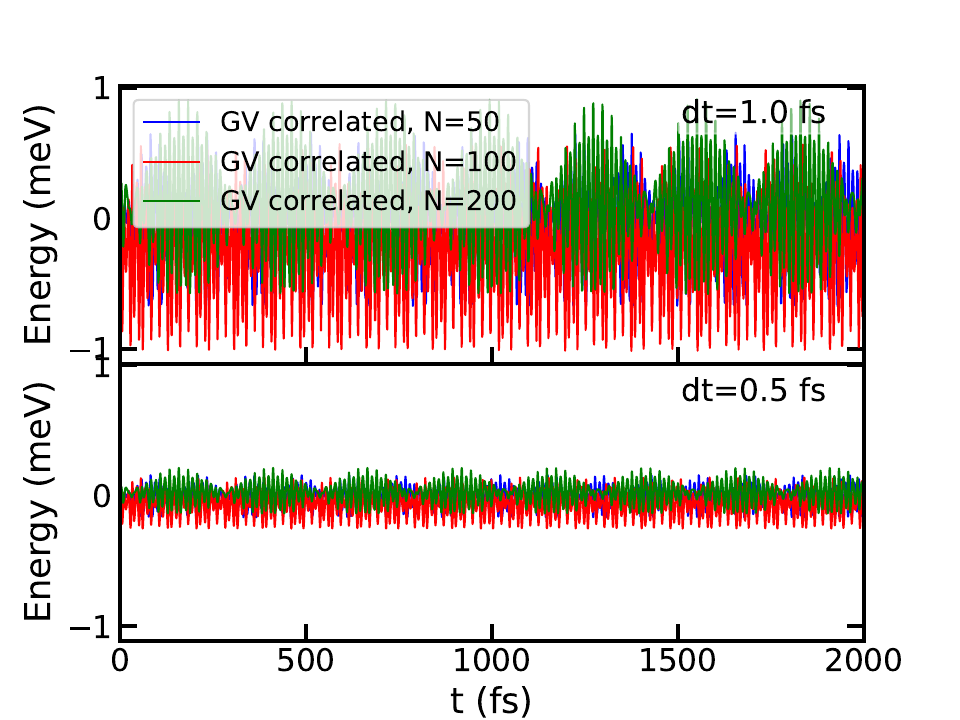}
        \caption{Correlated}
        \label{corr_ene}
    \end{subfigure}
    \caption{(a) Energy conservation of the GV method as a function of the number of configurations $N_c$ in the uncorrelated formulation. (b) Energy conservation in the correlated formulation. Note that the y-scale of the two panels differ by more than two orders of magnitudes.}
    \label{ene_conf}
\end{figure}

\section{Dynamics in SrTiO$_3$}\label{STO}
In this section, we showcase a realistic application of TD-SCHA by investigating the out-of-equilibrium quantum dynamics in \ch{SrTiO3} (STO) that follows a resonance-exciting short pulse of infrared light. STO is a prototypical quantum paraelectric \cite{PhysRevB.104.L060103, PhysRevMaterials.7.L030801, PhysRevResearch.4.033020}, where nuclear quantum fluctuations suppress the ferroelectric order at low temperatures. STO has been extensively studied due to the emerging phenomena occurring when driven out of equilibrium by strong electric field pulses \cite{doi:10.1126/science.aaw4911, doi:10.1126/science.aaw4913, Kozina2019, PhysRevLett.129.167401, PhysRevLett.108.097401, orenstein2024observation}. Notably, irradiating STO with a THz-frequency pulse at low temperatures induces a long-lasting second harmonic generation signal \cite{doi:10.1126/science.aaw4911, doi:10.1126/science.aaw4913}, suggesting the occurrence of a light-induced ferroelectric phase transition; however, this interpretation is still  debated\cite{PhysRevLett.130.126902}. Furthermore, recent studies have demonstrated the possibility of transferring energy from lower frequency phonons, pumped by the optical excitation, to higher frequency phonons in an out-of-equilibrium process called upconversion\cite{Kozina2019}, enabled by the anharmonic coupling between them.

In our simulation, a 40-atom supercell of STO originally equilibrated at \SI{100}{\kelvin} through a static SSCHA calculation is excited by an infrared pulse with an amplitude of 833 kV/cm, which is resonant with the soft phonon mode (SPM, represented in Fig. \ref{figure_sto}\textcolor{blue}{a} of STO. We account for the light-matter interaction in the dipole approximation through the Born effective charges. Details on the coupling with the electric field and the atomic energy landscape calculation are discussed in appendix \ref{ml_potential}. We integrate the TD-SCHA equations using the GV scheme in the correlated formulation, adopting a time step of 1 fs and sampling the potential energy landscape with $N_c$=4000.
The system, originally in equilibrium at \SI{100}{\kelvin}, interacts at $t=\SI{0}{fs}$ with an external pulse of oscillating electric field, triggering a non-equilibrium evolution of the density matrix. Fig. \ref{figure_sto}\textcolor{blue}{b} shows the motion of the SPM phonon coordinate 
\begin{equation}
    Q_{\mu} = \sum_{ax} e_{\mu ax}(\mathcal{R}_{ax} - \mathcal{R}_{ax}^{eq})
\end{equation}
as a function of the time delay after the pulse. Here $\mathcal{R}_{ax}^{eq}$ represents the equilibrium centroid position of the atom $a$ in the direction $x$, and $e_{\mu ax}$ is the equilibirum eigenvector of the soft phonon mode $\mu$. The irradiation of STO with resonant pulses drives large oscillations of the SPM, which slowly decay due to the interaction with other phonon modes. The blue area in the figure corresponds to the quantum uncertainty in the position of the SPM. It is computed as $\pm \sqrt{A_{\mu\mu}}$, which is equal to $\sqrt{\braket{Q_{\mu}^2}}$. The large extent of this uncertainty relative to the motion of $Q_{\mu}$ highlights the fundamental importance of quantum effects in the dynamics of STO.
An extensive discussion of the relevance of the simulation for the physics of STO goes beyond the scope of this work, and is subject of a separate publication \cite{lavoro_sto}.
\begin{figure}
    \centering
\begin{subfigure}{0.5\textwidth}
    \includegraphics[scale=0.3]{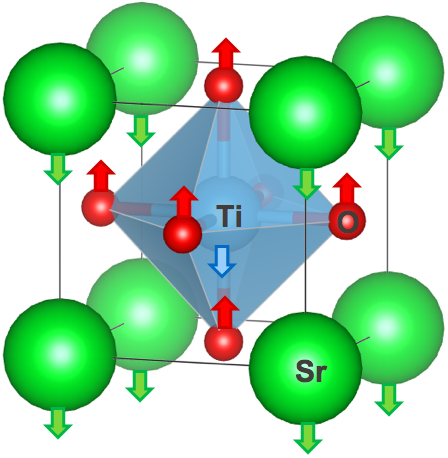}
    \caption{}
    \label{}
\end{subfigure}
\begin{subfigure}{0.5\textwidth}
    \includegraphics[scale=0.45]{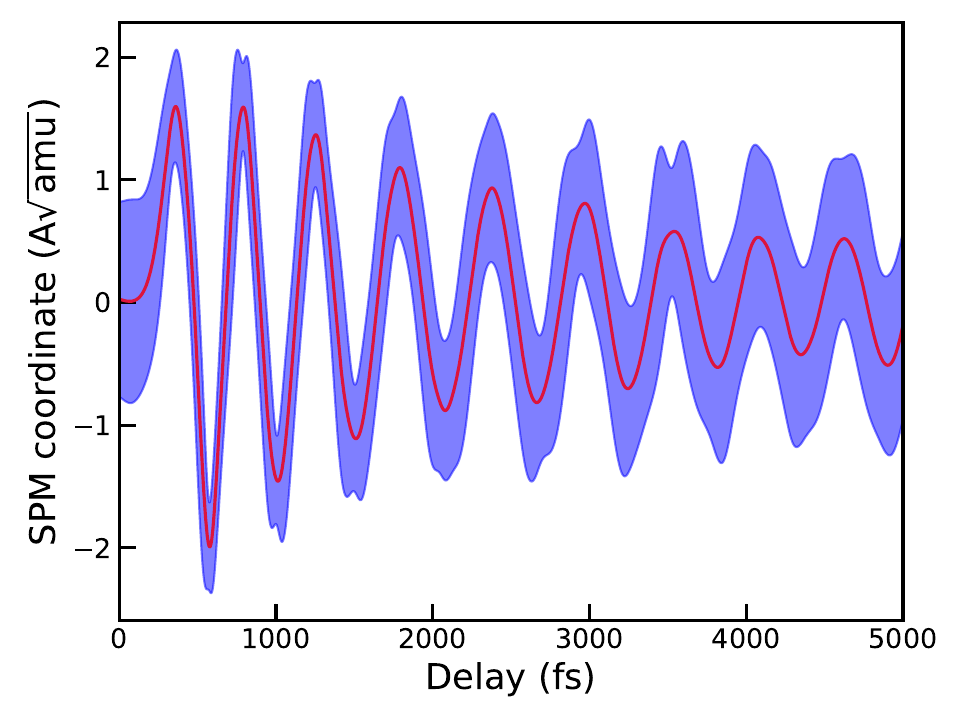}
    \caption{}
    \label{figure_sto}
\end{subfigure}
\caption{(a) The STO unit cell. The arrows indicate the displacement pattern of the SPM excited resonantly by the THz impulsive pump. (b) Dynamics of the SPM as a function of the time delay after the pulse. The blue area represents the quantum uncertainty, which is comparable to the amplitude of the oscillations, witnessing the importance of nuclear quantum effects in the dynamics. }
\label{figure_sto}
\end{figure}

\section{Conclusions}
In this work, we introduced the first approach to simulate nonequilibrium quantum nuclear dynamics using the TD-SCHA. We derived an integration scheme, the Generalized Verlet, which allows for the evolution of the equations with an error of $\mathcal{O}(dt^3)$, demonstrating that its conditional stability is consistent and of the same order as the Nyquist sampling rate. Additionally, we introduced a stochastic formulation of the TD-SCHA, which enables efficient calculation of ensemble averages while ensuring the stability of the evolution. Finally, we showcased the method's potential, proving it is well-suited for simulating quantum nonequilibrium processes in pump-probe setups on the scale of hundreds of atoms.

\section*{Acknowledgements}
This research was partially funded by the Swiss National Science Foundation (SNSF, mobility fellowship P500PT\_217861), the Department of Navy award N00014-20-1-2418 issued by the Office of Naval Research and Robert Bosch LLC.
L. M. thanks the European Union under the program Horizon 2020 for the award and funding of the MSCA individual fellowship (grant number 101018714). Computational resources were provided by the FAS Division of Science Research Computing Group at Harvard University. 

\section*{Contributions}
F.L. conducted the theoretical derivations, implemented the methods, and carried out the atomistic simulations. A.J. trained the machine-learned potential. L. M. guided the numerical implementation and algorithminc development. L.M. and B.K. supervised the full project.
\newpage

\newpage
\onecolumngrid
\appendix

\section{Notes on the Wigner formulation of TD-SCHA}\label{wigner}
The nuclear density introduced in Eq. \ref{density} corresponds to the most general Gaussian form in the positions $\bR$ and momenta $\bP$ variables. It is parametrized by the vectors $\mR$ and $\mP$ and the matrices $\bA$, $\bB$, and $\bG$. The physical meaning of these parameters is immediately clear after noting that
\begin{equation}\label{avgR}
    \mR(t) = \braket{\bR}(t)\ ,
\end{equation}
\begin{equation}\label{avgP}
    \mP(t) = \braket{\bP}(t) \ ,
\end{equation}
\begin{equation}\label{avgA}
    A_{ij}(t) = \braket{\delta R_i \delta R_j} \ ,
\end{equation}
\begin{equation}
    B_{ij}(t) = \braket{\delta P_i \delta P_j} \ ,
\end{equation}
\begin{equation}\label{avgG}
    \Gamma_{ij}(t) = \braket{\delta R_i \delta P_j}\ .
\end{equation}
These properties can be easily shown using the definition of ensemble averages given in Eq. \ref{avgO}.
According to Eqs. \ref{avgR} and \ref{avgP}, the parameters $\mR$ and $\mP$ correspond to the expected values for the positions and momenta, which are analogous to the classical positions and momenta. The variables $\bA$, $\bB$ and $\bG$, instead, correspond to the position-position, momentum-momentum and position-momentum correlation matrices, as suggested by Eqs. \ref{avgA}-\ref{avgG}.\\
If an observable is a function of only the position $O(\bR)$, the momenta degrees of freedom can be integrated out in Eq. \ref{avgO}. After some tedious algebra, it is possible to show that in such case 
\begin{equation}\label{avg_simple}
    \braket{O} = \int   O(\mR(t) + \delta \mathbf{R})\ \sqrt{\frac{1}{(2\pi)^{3N}\mathrm{det}\ A}} e^{-\frac{1}{2}\delta \mathbf{R\cdot} \bA^{-1}(t)\cdot \delta \mathbf{R}} \ d(\delta \mathbf{R})\ .
\end{equation}
We can clearly see that $\braket{O}(t)$ is a function of $\mR$ and $\bA$ only.

\section{Numerical Stability of Integation Schemes}\label{integr}
First of all, we prove an identity for the central finite difference approximation of a derivative. We first write the Taylor expansion of a function $F$ up to the third order centered at $t$:
\begin{equation}
    F_{t+dt} = F_t + F'_tdt + \frac{1}{2}F''_tdt^2 + \mathcal{O}(dt^3)
\end{equation}
 we can write the same expression by centering it in $t+dt$,
\begin{equation}
    F_t = F_{t+dt} - F'_{t+dt}dt + \frac{1}{2}F''_{t+dt}dt^2 + \mathcal{O}(dt^3)
\end{equation}
Adding up these equations we get
\begin{equation}
    F_{t+dt} = F_t + \frac{1}{2}(F'_t+F'_{t+dt})dt + \frac{1}{2}(F''_t-F''_{t+dt})dt^2 + \mathcal{O}(dt^3)\ .
\end{equation}
However, from the following equation 
\begin{equation}
    F''_{t+dt} = F''_t + F'''_tdt + \mathcal{O}(dt^2)
\end{equation}
we get
\begin{equation}
    (F''_{t+dt} - F''_t)dt^2 = F'''_tdt^3 + \mathcal{O}(dt^4)
\end{equation}
leading to the result
\begin{equation}
    F_{t+dt} = F_t + \frac{1}{2}(F'_t+F'_{t+dt})dt + \mathcal{O}(dt^3)\ ,
\end{equation}
which corresponds to Eq. \ref{mixed} of the main text. This equation indicates that we can evolve a function $F$ with accuracy up to third-order by utilizing the average of the derivative at time $t$ and that at time $t+dt$.
Eq. \ref{mixed} can be employed to derive the Verlet algorithm:
\begin{equation}
\begin{cases}
    \mP_{t+dt} = \mP_t + \frac{1}{2}\Bigl(\af_t+ \af_{t+dt}\Bigr)dt + \mathcal{O}(dt^3)   \\
        \mR_{t+dt} = \mR_t + \mP_tdt + \frac{1}{2}\af_t dt^2 + \mathcal{O}(dt^3) 
\end{cases}\ .
\end{equation}

In the following, we study the stability of the explicit Euler, semi-implicit Euler and generalized Verlet methods. 
\textbf{1. Explicit Euler method}\\
At each step of the explicit Euler method, the variables are updated according to the following rule:
\begin{equation}
    \mathbf{x}_{n+1} =  \begin{pmatrix}
        1 &  0 & dt\\
        0 &  1 & -dt\\
        -dt & dt & 1
    \end{pmatrix} \mathbf{x}_n = \mathbf{S}(dt)\mathbf{x}_n 
\end{equation}
leading to 
\begin{equation}\label{nterm}
    \mathbf{x}_{n+1}= \mathbf{S}(dt)^n \mathbf{x}_0  
\end{equation}
The exponentiation of such a matrix requires the calculation of its eigenvalue, which are
\begin{equation}
    \begin{cases}
        \lambda_1 = 1 \\
        \lambda_2 = 1 - i\sqrt{2}dt\\
        \lambda_3 = 1 + i\sqrt{2}dt \\
    \end{cases}
\end{equation}
The stability condition for preventing divergence of the power sequence in Eq. \ref{nterm} is
\begin{equation}
    |1\pm i\sqrt{2}dt|\leq1
\end{equation}
which is never satisfied.  The explicit Euler method is thus \textit{unconditionally unstable}.
\\\\
\textbf{2. Semi-implicit Euler method}\\
At each step, first $\Gamma$ is updated:
\begin{equation}\label{gamma1}
    \Gamma_{n+1} = \Gamma_{n} + (B_n-A_n)dt
\end{equation}
and then $A$ and $B$
\begin{equation}\label{a1}
\begin{cases}
        A_{n+1} = A_{n} + \Gamma_{n+1}dt \\
        B_{n+1}   = B_{n} - \Gamma_{n+1}dt
\end{cases}
\end{equation}
Substituting Eq. \ref{gamma1} into  \ref{a1}, we obtain
\begin{equation}
 \begin{cases}
        A_{n+1} = A_{n} + \Gamma_{n}dt + (B_n -A_n)dt^2\\
        B_{n+1}   = B_{n} - \Gamma_{n}dt - (B_n-A_n)dt^2
\end{cases}   
\end{equation}
or equivalently
\begin{equation}
    \begin{pmatrix}
        A_{n+1}\\ B_{n+1}\\ \Gamma_{n+1}
    \end{pmatrix}
    = 
    \begin{pmatrix}
        1-dt^2 & dt^2 & dt\\
        dt^2 & 1-dt^2 & -dt\\
        -dt & dt & 1
    \end{pmatrix}
    \begin{pmatrix}
        A_n \\ B_n \\ \Gamma_n
    \end{pmatrix}\ .
\end{equation}
The eigenvalues of the step matrix are
\begin{equation}
    \begin{cases}
        \lambda_1 = 1 \\
        \lambda_2 = 1 - dt^2 -dt\sqrt{dt^2-2}\\
        \lambda_3 = 1 - dt^2 +dt\sqrt{dt^2-2} \\        
    \end{cases}
\end{equation}
The eigenvalues $\lambda_2$, $\lambda_3$ are real for $dt\geq\sqrt{2}$, or complex conjugated otherwise. We easily note that
\begin{equation}
    \lambda_2 <-1 \qquad \forall dt \geq \sqrt{2}
\end{equation}
meaning that this method is \textit{unstable} in such range. 
For $dt<\sqrt{2}$, $\lambda_{2,3}$ are complex, with modulus
\begin{equation}
    |\lambda_{2,3}|^2 = (1-dt^2)^2 + 2dt^2-dt^4 = 1
\end{equation}
therefore the method is \textit{stable}.
Remembering the definition \ref{transf}, the stability condition for this method is
\begin{equation}
    t\leq \frac{1}{\sqrt{\kappa}} = \frac{1}{\omega}
\end{equation}
where $\omega$ is the frequency of the harmonic oscillator (we remind that $\kappa = \frac{1}{m}\frac{\partial^2V}{\partial R^2}$ due to the mass rescaling convention adopted). 
\\\\
\textbf{3. Generalized Verlet}\\
The same analylisis can be performed on the GV algorithm. 
In this case, the variables are updated according to 
\begin{equation}
    \begin{cases}
        A_{n+1} = A_n + \Gamma_n dt + \frac{1}{2}(B_n- A_n)dt^2 \\
        B_{n+1} = B_n - \frac{1}{2}(\Gamma_n + \Gamma_{n+1})dt\\
        \Gamma_{n+1} = \Gamma_n + \frac{1}{2}(B_n - A_n + B_{n+1}-A_{n+1} )dt
    \end{cases}
\end{equation}
After some tedious algebra, it is possible to write explicitly the transformation in matrix form as

\begin{equation}
    \begin{pmatrix}
        A_{n+1}\\ B_{n+1}\\ \Gamma_{n+1}
    \end{pmatrix}
    = 
    \begin{pmatrix}
        1-\frac{dt^2}{2} & \frac{dt^2}{2} & dt\\
        \frac{-dt^4+4dt^2}{2dt^2+8} & \frac{dt^4-2dt^2+8}{2dt^2+8} & \frac{dt^3-4dt}{dt^2+4}\\
        \frac{dt^3-4dt}{dt^2+4} & \frac{-dt^3+4dt}{dt^2+4}  & \frac{-3dt^2+4}{dt^2+4} 
    \end{pmatrix}
    \begin{pmatrix}
        A_n \\ B_n \\ \Gamma_n
    \end{pmatrix}\ ,
\end{equation}
with eigenvalues 
\begin{equation}
    \begin{cases}
        \lambda_1 = 1 \\
        \lambda_2 = \frac{4 - 3dt^2 -2\sqrt{2}dt\sqrt{dt^2-4}}{dt^2+4}\\
        \lambda_3 = \frac{4 - 3dt^2 +2\sqrt{2}dt\sqrt{dt^2-4}}{dt^2+4} \\        
    \end{cases} \ . 
\end{equation}
The eigenvalues are complex for $dt<2$, with modulus
\begin{equation}
    |\lambda_{2,3}|^2 = 1 \ ,
\end{equation}
and the method is stable, whereas for $dt>2$ 
\begin{equation}
    \lambda_2 < -1 \ ,
\end{equation}
thus the method is unstable.
To conclude, the stability condition for such an algorithm is
\begin{equation}
    dt < \sqrt{\frac{2}{\kappa}} = \frac{\sqrt{2}}{\omega}\ .
\end{equation}
The stability range is thus larger than that of the SIE.

\section{Importance sampling}\label{imp_samp}
In the following section we define some properties which will be useful for the development of the theory reported in the main text. 
The ensemble average of the potential energy at time t is given by
\begin{equation}\label{avg}
    \braket{V} = \int   V(\mR(t) + \delta \mathbf{R})\ \sqrt{\frac{1}{(2\pi)^{3N}\mathrm{det}\ A}} e^{-\frac{1}{2}\delta \mathbf{R\cdot} \bA^{-1}(t)\cdot \delta \mathbf{R}} \ d(\delta \mathbf{R})
\end{equation}
where the integration is carried over the variables $\delta R_a$ whose number is $3N$. 
We now perform a change of variables which is meant to turn the Gaussian in Eq. \ref{avg} into a normal distribution. We start from the modal decomposition of the matrix $A$
\begin{equation}
    A_{ab} = \sum_{\mu}\lambda_{\mu}e_{\mu a}e_{\mu b}\ ,
\end{equation}
and express the displacement $\delta R_a$ in normal coordinates:
\begin{equation}\label{nc1}
    \delta R_{a} = \sum_{b} J_{ab} y_{b} \ .
\end{equation}
The determinant of the Jacobian matrix is 
\begin{equation}
    \mathrm{det}\ J = \prod_{\mu} \sqrt{\lambda_{\mu}}  = \sqrt{ \mathrm{det}\ A}\ .
\end{equation}
The change of variables of Eq. \ref{nc1} turns the integral \ref{avg} into
\begin{equation}\label{avg1}
    \braket{V} = \int V(\mathcal{R}_a(t) + \sum_{b}J_{ab}(t)y_{b})\ \prod_{b}\frac{e^{-\frac{1}{2}y_{b}^2 }}{\sqrt{2\pi}}dy_{b}\ .
\end{equation}
Eq. \ref{avg1} provides the starting point for the stochastic calculation of the ensemble averages, expressed in Eq. \ref{vD} of the main text.

\section{Energy conservation}\label{energy_conservation}
Here, we outline the main steps introduced in Ref. \cite{PhysRevB.107.174307} to derive energy conservation for infinitely many configurations. We then show the conditions under which such conservation holds regardless of the number of configurations, which represents the result of the current work. The time derivative of the quantum kinetic energy reads
\begin{equation}\label{dke}
    \frac{d}{dt} \sum_i \Bigl\langle \frac{P_i^2}{2} \Bigl\rangle = \frac{d}{dt} \frac{\sum_i\mathcal{P}_i^2 + \mathrm{Tr}\ \bB}{2} = 
    \sum_i \Bigl( \mathcal{P}_i\dot{\mathcal{P}}_i + \frac{\dot{B}_{ii}}{2} \Bigr)
\end{equation}
Combining the first and second of Eqs. \ref{tdscha}, we obtain
\begin{equation}
    \sum_i \mathcal{P}_i\dot{\mathcal{P}}_i = -\sum_i \Bigl\langle\frac{\partial V}{\partial R_i}\Bigr\rangle \dot{\mathcal{R}}_i\ , 
\end{equation}
while the third and fourth of Eqs. \ref{tdscha} give
\begin{equation}
    \sum_i \frac{\dot{B}_{ii}}{2} = -\frac{1}{2}\sum_{ij} \Bigl\langle\frac{\partial^2 V}{\partial R_i \partial R_j}\Bigr\rangle \dot{A}_{ij}\ . 
\end{equation}
Importantly, these equations also hold when the discrete expressions for forces (Eq. \ref{fD}) and curvatures (Eq. \ref{kDs}) are used to drive the dynamics
\begin{equation}
    \sum_i \mathcal{P}_i\dot{\mathcal{P}}_i = -\sum_i \Bigl\langle\frac{\partial V}{\partial R_i}\Bigr\rangle_{\mathcal{D}} \dot{\mathcal{R}}_i\ , 
\end{equation}
\begin{equation}
    \sum_i \frac{\dot{B}_{ii}}{2} = -\frac{1}{2} \sum_{ij} \Bigl\langle\frac{\partial^2 V}{\partial R_i \partial R_j}\Bigr\rangle_{\mathcal{D}} \dot{A}_{ij}\ . 
\end{equation}
Following Ref. \cite{PhysRevB.107.174307}, it is possible to show that
\begin{equation}\label{dr}
   \Bigl\langle\frac{\partial V}{\partial R_i}\Bigr\rangle = \frac{\partial \braket{V}}{\partial \mathcal{R}_i} 
\end{equation}
and 
\begin{equation}\label{da}
   \frac{1}{2}\Bigl\langle\frac{\partial^2 V}{\partial R_iR_j}\Bigr\rangle = \frac{\partial \braket{V}}{\partial \mathcal{A}_{ij}} \ .
\end{equation}
Inserting Eqs. \ref{dr} and \ref{da} into Eq. \ref{dke} we obtain
\begin{equation}
    \frac{1}{2}\frac{d}{dt}\Bigl( {\sum_i\mathcal{P}_i^2 + \mathrm{Tr}\ \bB}\Bigr) = 
    - \frac{\partial \braket{V}}{\partial \mathcal{R}_i} \dot{\mathcal{R}}_i
    - \frac{\partial \braket{V}}{\partial \mathcal{A}_{ij}} \dot{A}_{ij} = -\frac{d}{dt}\braket{V}\ ,
\end{equation}
This proves the conservation of energy in absence of external forces. However, for this energy conservation to hold for any configuration, we need to demonstrate that Eqs. \ref{dr} and \ref{da} also hold for the discrete expression of forces and curvatures, namely:
\begin{equation}\label{drd}
   \Bigl\langle\frac{\partial V}{\partial R_i}\Bigr\rangle_{\mathcal{D}} = \frac{\partial \braket{V}_{\mathcal{D}}}{\partial \mathcal{R}_i} 
\end{equation}
and 
\begin{equation}\label{dad}
   \frac{1}{2}\Bigl\langle\frac{\partial^2 V}{\partial R_iR_j}\Bigr\rangle_{\mathcal{D}} = \frac{\partial \braket{V}_{\mathcal{D}}}{\partial \mathcal{A}_{ij}} \ .
\end{equation}
Equation \ref{drd} can be easily derived by applying the chain rule to differentiate Equation \ref{vD}:
\begin{equation}
        \Bigl\langle \frac{\partial V}{\partial R_a}\Bigr\rangle_{\mathcal{D}} = -\frac{1}{N_c}\sum_{i=1}^{N_c} f_a(\mR + \mathbf{J}\cdot\mathbf{y}_i) = \frac{1}{N_c}\sum_{i=1}^{N_c} \frac{\partial V}{\partial \mathcal{R}_a}(\mR + \mathbf{J}\cdot\mathbf{y}_i) = \frac{\partial \braket{V}_{\mathcal{D}}}{\partial \mathcal{R}_a} \ ,  
\end{equation}
The derivation of Eq. \ref{dad} is more complex, and does not hold in the most general case. From Eqs. \ref{kD} and \ref{kDs}, we know that 
\begin{equation}
        \Bigl\langle \frac{\partial^2 V}{\partial R_a \partial R_b}\Bigr\rangle_{\mathcal{D}}^{sym} = -\frac{1}{2}\sum_{cd} A^{-1}_{ac}\sum_{i=1}^N J_{cd}y_{d i} f_b(\mR + \mathbf{J}\cdot\mathbf{y}_i)\  
        -\frac{1}{2}\sum_{cd} A^{-1}_{bc}\sum_{i=1}^N J_{cd}y_{d i} f_a(\mR + \mathbf{J}\cdot\mathbf{y}_i)\ .
\end{equation}
At the same time, the derivative of $\braket{V}_{\mathcal{D}}$ with the respect to the element $A_{ab}$ is
\begin{equation}\label{dda}
    \frac{\partial }{\partial A_{ab}}\braket{V}_{\mathcal{D}} = 
    -\frac{1}{N_c}\sum_{cd}\sum_{i=1}^{N_c} f_c(\mR + \mathbf{J}\cdot\mathbf{y}_i) \frac{\partial J_{cd}}{\partial A_{ab}}y_d
\end{equation}
In order to prove Eq. \ref{dad}, it would be enough to show that 
\begin{equation}\label{prob}
    \frac{\partial J_{cd}}{\partial A_{ab}} = \frac{1}{4}\delta_{ac} J^{-1}_{bd} + \frac{1}{4}\delta_{bc} J^{-1}_{ad} \ . 
\end{equation}
In one dimensional problems, identity \ref{prob} reduces to 
\begin{equation}
    \frac{\partial J}{\partial A} = \frac{1}{2\sqrt{A}}\ , 
\end{equation}
which is satisfied since $J=\sqrt{A}$. Therefore, in one dimensional problems, the conservation of energy holds independently on the number of configurations, provided that the time step is reasonably small.  \\
For higher dimensional problems, the validity of the identity $\ref{prob}$ depends on the definition of $\mathrm{\mathbf{J}}$. In fact, there are infinite many ways to obtain the square root of a symmetric matrix, defined as the matrix $\mathrm{\mathbf{J}}$ such that 
\begin{equation}\label{sqrt}
    \mathrm{ \mathbf{A} = \mathbf{J} \mathbf{J}^T } \ .
\end{equation}
This is because multiplying the matrix $\mathrm{\mathbf{J}}$ by any orthogonal matrix $\mathbf{O}$ results in a matrix that still satisfies Eq. \ref{sqrt}.
The definition of  $\mathrm{\mathbf{J}}$ in Eq.  \ref{J} does not satisfy the identity \ref{prob}. In fact, using the rules derived in Appendix 5 of Ref. \cite{PhysRevB.96.014111}, we easily obtain
\begin{equation}
    \sum_{ab}\frac{\partial J_{cd}}{\partial A_{ab}}\dot{A}_{ab} = \sum_{ab}\sum_{\mu\nu}
    \frac{e_{a\mu}e_{b\nu}e_{c\mu}e_{d\nu}}{\sqrt{\lambda_{\mu}} + \sqrt{\lambda_{\nu}}}  \dot{A}_{ab} \ .
\end{equation}
The problem of finding an expression for $\mathrm{\mathbf{J}}$ that satisfies identity \ref{prob} is very important and will be the subject of future research efforts.

\section{Atomic energy landscape}\label{ml_potential}
The TD-SCHA simulations on STO utilize machine-learned potentials (MLP) to model the potential energy surface. We opted for FLARE~\cite{Vandermause2020} for its active learning capabilities, enabling efficient data generation and rapid inference times~\cite{trill}. Through active learning, we explored various temperatures and volumes to develop a broadly applicable potential.\\
Simulations were conducted at 100, 300, and 500 K, each for 200 ps, at the DFT-relaxed lattice parameter, as well as at $\pm$2\% strain, amounting to a total of 1.8 ns of dynamics. A timestep of 2 fs and a thermostat damping time of 200 fs were employed, utilizing the default Nosé–Hoover thermostat in LAMMPS~\cite{lammps22}.\\
DFT calculations were performed at the PBE level of theory using Quantum ESPRESSO \cite{Giannozzi_2009}. We used a plane-wave cutoff of 80 Ry and a k-point grid of $\mathrm{6\times6\times8}$ for the 20-atom cell, adopting the pseudopotentials recommended by the SSSP efficiency library \cite{prandini2018precision}.\\
The forces on the atom $i$ by the electric field is obtained as
\begin{equation}
    \mathbf{f}_i= \frac{1}{\varepsilon_{eff}}\ \mathbf{Z}_i\cdot \mathbf{\mathcal{E}}\ .
\end{equation}
Here $\mathbf{\mathcal{E}}$ is the external electric field,  $\varepsilon_{eff}$ is the dielectric constant and $\mathbf{Z}_i$ are the Born effective charge tensors. The effective charges are computed through DFPT, using the same parameters as above. Their value for the different atomic species is reported in Tab. \ref{Zeff}.
\begin{table}[h]
    \centering
    \begin{tabular}{c|c|c|c|}
           & $\mathrm{Z^*_{xx}}$ & $\mathrm{Z^*_{yy}}$ & $\mathrm{Z^*_{zz}}$\\ \hline\hline
        Ti & 7.338 & 7.338 & 7.338\\
        Sr & 2.549 & 2.549 & 2.549\\
        $\mathrm{O_1}$ & -2.024 & -5.845 & -2.024 \\
        $\mathrm{O_2}$ & -2.024 & -2.024&  -5.845\\
        $\mathrm{O_3}$ & -5.845 & -2.024 & -2.024\\
    \end{tabular}
    \caption{Born effective charges for the cubic STO unitcell, computed through DFPT.}
    \label{Zeff}
\end{table}
Here we employ the screening model proposed in Refs. \cite{PhysRevB.85.045134, PhysRevLett.129.167401} 
\begin{equation}
    \varepsilon_{eff} = \frac{1+\sqrt{\varepsilon_{DFPT}}}{2}\ ,
\end{equation}
with $\varepsilon_{DFPT}=6.31$.

%\bibliographystyle{ieeetr}
%\bibliography{main.bib}

%

\end{document}